\documentclass[twocolumn,aps,amssymb,,superscriptaddress]{revtex4}
\usepackage{graphicx}
\def\cm{cm$^{-1}$}

\begin{document}
\title{Magnetic and magnetoelectric excitations in TbMnO$_3$
}
\author{A. Pimenov}
\author{A. Shuvaev}
\affiliation{Experimentelle Physik IV, Universit\"{a}t W\"{u}rzburg,
97074 W\"{u}rzburg, Germany} %
\author{ A. Loidl}
\author{ F. Schrettle}
\affiliation{EP V, Center for Electronic Correlations and Magnetism,
University of Augsburg,
86135 Augsburg, Germany} %
\author{A.$\>$A. Mukhin}
\author{V.D. Travkin}
\author{V.Yu. Ivanov}
\affiliation{General Physics Institute of the Russian Acad. of
Sciences, 119991 Moscow, Russia}%
\author{A. M. Balbashov}
\affiliation{Moscow Power Engineering Institute, 105835 Moscow,
Russia}
\date{\today}

\begin{abstract}
Magnetic and magnetoelectric excitations in the multiferroic
TbMnO$_3$ have been investigated at terahertz frequencies. Using
different experimental geometries we can clearly separate the
electro-active excitations (electromagnons) from the magneto-active
modes, i.e. antiferromagnetic resonances (AFMR). Two AFMR resonances
were found to coincide with electromagnons. This indicates that both
excitations belong to the same mode and the electromagnons can be
excited by magnetic ac-field as well. In external magnetic fields
and at low temperatures distinct fine structure of the
electromagnons appears. In spite of the $90^o$ rotation of the
magnetic structure, the electromagnons are observable for  electric
ac-fields parallel to the a-axis only. Contrary to simple
expectations, the response along the c-axis remains purely magnetic
in nature.

\end{abstract}


\maketitle

During the last years, materials with magnetoelectric (ME) coupling
have attracted much interest especially due to their intriguing
physical mechanisms and their potential for applications
\cite{fiebig,tokura,cheong,khomskii,ramesh}. ME coupling is
especially strong in multiferroics, i.e materials which are
simultaneously ferroelectric and ferromagnetic. The strength of ME
effects in multiferroics is due to direct coupling of the magnetic
and electric order parameters and partly due to improper character
of the ferroelectricity. This coupling allows, for example,
switching of electric polarization in the sample in external
magnetic fields \cite{kimura,goto}. One promising class of
multiferroics is represented by frustrated magnets \cite{cheong} in
which magnetoelectricity is induced by complex spin arrangements
like cycloidal or spiral antiferromagnetic structures.

Given the observation of the static magnetoelectric effects in
susceptibilities and polarizations, the existence of the dynamic
effects can be expected from the first principles. This follows
immediately from the optical sum rules which arises as a direct
consequence of the causality. Indeed, strong magnetoelectric modes
have been observed in multiferroic manganites TbMnO$_3$ and
GdMnO$_3$ and termed electromagnons \cite{nphys}. The electromagnons
have been detected in the infrared spectra of YMn$_2$O$_5$,
TbMn$_2$O$_5$ \cite{sushkov},  (Eu:Y)MnO$_3$ \cite{aguilar,euymno},
and DyMnO$_3$ \cite{kida} using similar experimental techniques.
Soft magnon modes observed in the inelastic neutron scattering data
of TbMnO$_3$ has been attributed to electromagnons as well
\cite{braden,braden2}. Based on available experimental data and on
existing theoretical models, electromagnons can be defined as spin
modes which become excited by electric field due to ME coupling
\cite{nphys,review}. Large spectral weight of the electromagnons
distinguishes them \cite{review} from seignetomagnons in ME
compounds, as predicted about forty years ago by Baryachtar and
Chupis \cite{chupis,chupis2}.

In spite of an enormous progress in the field of ME effect the
underlying microscopic mechanisms still remain under debate. In
order to explain spin driven ferroelectricity and dynamical
properties of frustrated magnets various approaches like
phenomenological analysis \cite{mostovoy,harris}, Brillouin-zone
folding \cite{sousa}, spin current model \cite{katsura05,katsura},
inverse Dzyaloshinskii-Moriya (DM) interaction \cite{sergienko} and
direct Heisenberg exchange \cite{sergienko2,sushkov2} have been
suggested. Recently it has been suspected that some of the
excitations detected in far infrared and terahertz (THz)
spectroscopy might be due to two-magnon scattering processes
\cite{kida,kida2,kida3}.

Among magnetoelectric manganites, TbMnO$_3$ is probably one of the
most intensively studied using spectroscopic techniques. In addition
to results from dielectric \cite{kimura,kimura05} and optical
\cite{nphys,kida2} spectroscopies, the magnetic structure of this
material is well known from neutron scattering
\cite{quezel,kajimoto,kenzelmann} and x-ray
\cite{argyriou,strempfer} experiments. In addition, inelastic
neutron scattering data are available for TbMnO$_3$
\cite{braden,braden2,braden3} which allow to compare characteristic
frequencies of spin excitations and of electromagnons. At $T_N =
42$\,K TbMnO$_3$ orders antiferromagnetically with the magnetic
moments of manganese ions aligned along the b-axis with an
incommensurate sinusoidal modulation
\cite{quezel,kajimoto,kenzelmann}. Upon cooling a second transition
into a cycloidal (spiral) phase occurs at $T_C = 28$\,K with a
slightly different modulation vector \cite{kenzelmann}. This low
temperature phase is ferroelectric with spontaneous polarization
parallel to the c-axis \cite{kimura}. Finally, a phase transition at
about 9\,K is attributed to the magnetic ordering of the Tb
sublattice. This ordering only weakly affects the ferroelectric and
dielectric properties of TbMnO$_3$ \cite{kimura05}.

In this paper we present detailed investigations of the terahertz
properties of TbMnO$_3$ for different experimental geometries. This
allowed us to separate magnetic and magnetoelectric contributions in
the experimental spectra. The coincidence of  the characteristic
frequencies for pairs of these excitations indicates that
electromagnons can be excited by magnetic ac-field as well. Finally
we prove the c-axis in TbMnO$_3$ remains electrically silent
independent of the orientation of the magnetic cycloid.

Single crystals of TbMnO$_3$ have been grown using the floating-zone
method with radiation heating. Samples characterization using X-ray,
magnetic and dielectric measurements shoved an agreement with the
published results \cite{kimura05}. The transmittance experiments at
terahertz frequencies (3 cm$^{-1}$ $<\nu<$ 40 cm$^{-1}$) were
carried out in a Mach-Zehnder interferometer arrangement
\cite{volkov,lsmo} which allows measurements of amplitude and phase
shift in a geometry with controlled polarization of the radiation.
The absolute values of the complex dielectric permittivity
$\varepsilon^*=\varepsilon_1+i\varepsilon_2$ were determined
directly from the measured spectra using the Fresnel optical
formulas for the complex transmission coefficient. The experiments
in external magnetic fields up to 7 T were performed in a
superconducting split-coil magnet with polypropylene windows.

Carrying out transmittance experiment with different experimental
geometries and varying the polarization of the incident radiation in
most cases allows to separate unambiguously magnetic and dielectric
contributions to the measured spectra. However, in some cases both
contributions are equally strong \cite{lsmo} and four independent
experiments are necessary to extract both, the dielectric
permittivity $\varepsilon^* = \varepsilon_1 +i\varepsilon_2$ and the
magnetic permeability $\mu^*=\mu_1+i\mu_2$. In such cases and within
a good approximation the complex refractive index $n+i\kappa=
\sqrt{\varepsilon^* \mu^*}$ is the best representation to analyze
the observed results.

\begin{figure}[]
\includegraphics[width=6cm,clip]{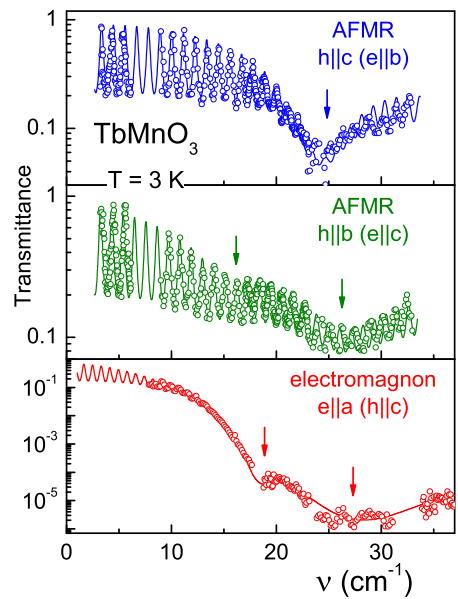}
\caption{(color online) Examples of terahertz transmittance spectra
of TbMnO$_3$ for different experimental geometries. Upper and middle
panels: antiferromagnetic resonance modes with excitation conditions
$\tilde{h}\|c$ and $\tilde{h}\|b$, respectively. Lower panel:
transmittance for a geometry $\tilde{e}\|a$ with electromagnons at
18 \cm and 26 \cm . Much lower transmittance in this case is due to
stronger intensity of electromagnons compared to AFMR. Symbols -
experiment, lines - fits using Lorentzian line shape. The
oscillations in the spectra are due to Fabry-P\'{e}rot interferences
on the sample surfaces. Specific geometry of each transmittance
experiment is given in parentheses.} \label{ftranb0}
\end{figure}

Figure \ref{ftranb0} shows three examples of the experimental
transmittance  of TbMnO$_3$ for different geometries. Upper and
middle panel of Fig. \ref{ftranb0} have been obtained in geometries
where magnetically excited modes are observed. We assign these modes
to antiferromagnetic resonances in TbMnO$_3$. Due to the comparative
weakness of these modes, the transmittance is not far from unity
even close to the resonance and the Fabry-P\'{e}rot oscillations on
the sample surfaces are clearly seen. On the contrary, the
excitation observed in the lover panel of Fig. \ref{ftranb0} reveals
much stronger absorption, which is partly close to the the
sensitivity limit of our spectrometer. As has been discussed
previously \cite{nphys,review}, these modes are excited by the
electric field and are termed electromagnons.

\begin{figure}[]
\includegraphics[width=6cm,clip]{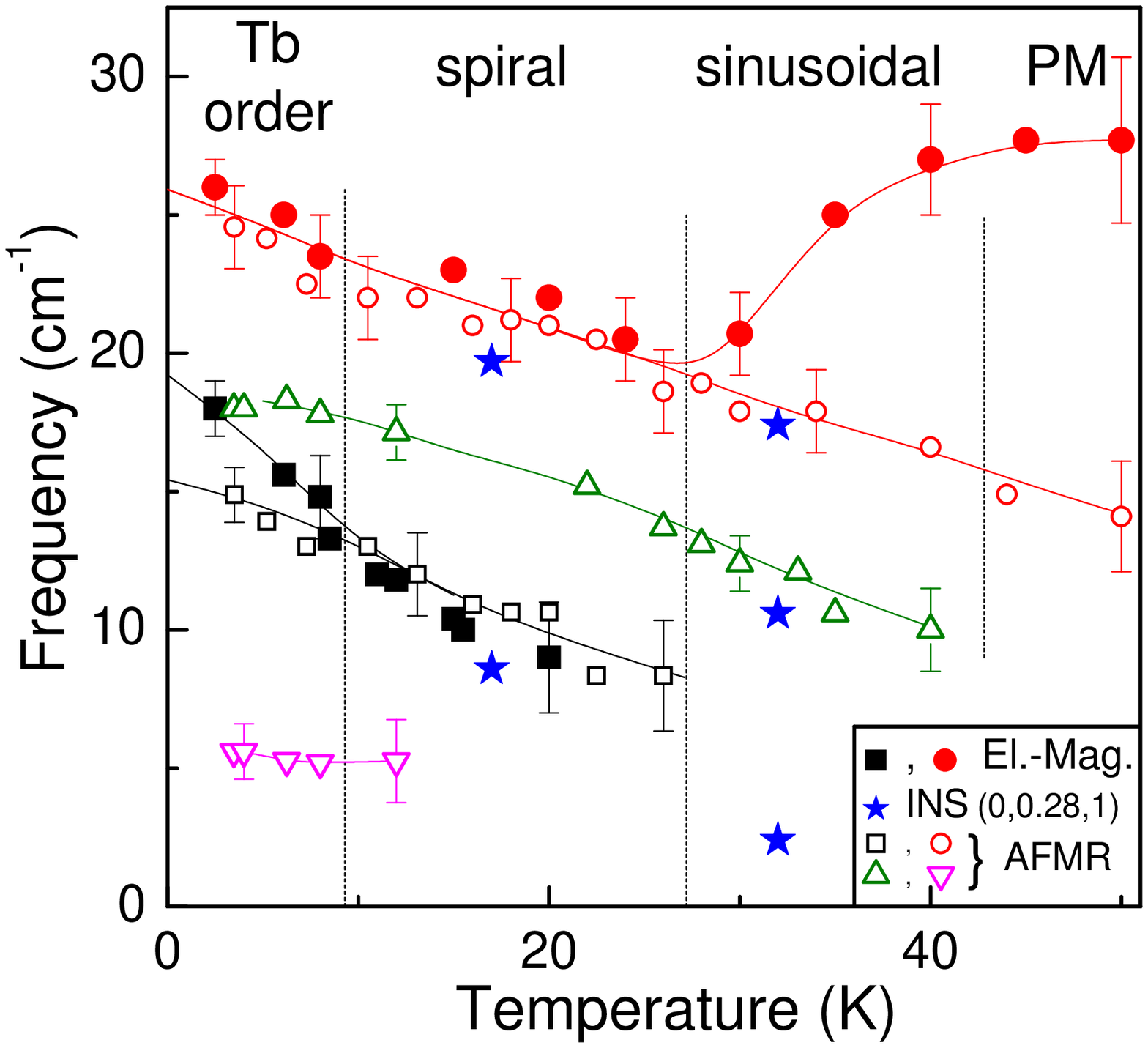}
\caption{(color online) Temperature dependence of the mode
frequencies of various excitations in TbMnO$_3$.  Solid symbols:
electromagnons which are observed for $\tilde{e}\|a$ only. In the
sinusoidal phase ($T>27$\,K) the electromagnons frequencies are not
well defined and the positions of the maxima in $\varepsilon_2$ are
plotted. Stars indicate the results from inelastic neutron
scattering experiments \cite{braden,braden2}. Open symbols:
antiferromagnetic resonances with the following excitation
conditions: circles $\tilde{h}\|b$ and $\tilde{h}\| c$, squares
$\tilde{h}\|b$, triangles $\tilde{h}\|a$. Lines are guide to the
eye.} \label{fmodes}
\end{figure}

As demonstrated in the lower panel of Fig. \ref{ftranb0} the
electromagnon mode splits into two excitations, which is most
clearly seen in the spectra at the temperatures below 4\,K.  Using
direct analysis of the transmittance spectra in combination with the
temperature scans, the second weaker electromagnon at 18 \cm can be
followed up to $T = 20$\,K, i.e. deep into the cycloidal phase. This
allowed us to compare the terahertz data with the results by
inelastic neutron scattering (INS) \cite{braden,braden2} where two
magnon modes have been observed in the cycloidal phase. In addition
to a high frequency excitation seen by both spectroscopies around 20
\cm, another mode has been observed close to 9 \cm. In our low
temperature data this second mode can be observed as well.

The frequencies of various excitations in TbMnO$_3$ are summarized
in Fig. \ref{fmodes}. In this figure two observed electromagnons are
indicated by solid circles and squares. In the spin spiral phase the
electromagnon energies correspond to the excitation energies of well
defined quasi particles. In the collinear sinusoidal phase the
electromagnons are seen broad over damped modes\cite{review}. The
energies as plotted for $T>27$\,K correspond to the line width of
these modes indicating that the damping strongly increases towards
the transitions into the paramagnetic phase.

The frequencies of the observed antiferromagnetic resonances in
TbMnO$_3$ are plotted in Fig. \ref{fmodes} as open symbols. In a
total, four such magnetic modes have been observed in the frequency
range of our experiment. Remarkably, both AFMR and electromagnons
can still be observed in the paramagnetic phase. This effect have
been previously observed for other multiferroics \cite{review} and
should be probably attributed to magnetic fluctuations.

One important result in Fig. \ref{fmodes} is the close coincidence
of two AFMR modes with electromagnons. In analogy to electromagnons,
these two modes are indicated by open circles (high frequency mode,
excited by $\tilde{h} \| b$ and $\tilde{h} \| c$) and open squares
(low frequency mode, excited by $\tilde{h} \| b$). Another AFMR mode
at intermediate frequencies which is given by open triangles can be
attributed to the phason mode of the magnetic bc-cycloid. This
agrees with the excitations conditions $\tilde{h} \| a$ for this
mode. The remaining magnetic mode around 5 \cm can be excited for
$\tilde{h} \| a$. Based on the fact that this mode is observed
mainly in the Tb-ordered phase, we attribute this mode to the
excitation of the ordered Tb moments.

In Fig. \ref{fmodes} the modes seen with INS technique
\cite{braden,braden2} are indicated by stars. Rough coincidence of
both excitations suggests that both spectroscopic techniques probe
the same mode. We recall however that the INS frequencies have been
obtained at nonzero wave-vector with $k_0\simeq 0.28$. On the
contrary, in the optical spectroscopy the relevant wave-vector
equals to the wave-vector of the photon and is always close to zero.
Therefore, this region of the magnon branch cannot be excited
directly and further mechanisms should play a role. In the present
case the apparent wave-vector contradiction is resolved due to
static modulation of the magnetic structure with the same
wave-vector $k_m=k_0\simeq 0.28$. In the presence of a periodic
modulation the \textit{umklapp} processes with $k_m$ become allowed
and the momentum conservation during the absorption of a photon with
$k_{ph} \approx 0$ can be fulfilled: $k_{ph}=k_m - k_0 \approx 0 $.
There still remain a controversy concerning possible assignment of
the INS frequencies to electromagnons or AFMR, which should be
resolved in future experiments.

On the basis of the polarization analysis of the spectra in
multiferroic manganites the electromagnons have been classified as
inhomogeneous spin modes which become electrically active due to ME
coupling \cite{review}. Due to magnetic origin of these modes it can
be also expected that the same modes will be excited by magnetic ac
field as well. This explains the observed coincidence of two
electromagnons with the AFMR modes. Full explanation of the observed
modes including excitations conditions is still lacking, which is
probably due to the complexity of the magnetic structure. Our
preliminary theoretical analysis of the spin oscillations in the
bc-cycloidal phase shows the existence of spin modes of two
different types. Firstly, one expect the existence of the modes
which can be excited both by magnetic field $\tilde{h}\|b,c$-axes
and by electric field $\tilde{e}\|a$-axis. In the simple model
without magnetic anisotropy within the bc-plane, these modes are
twice degenerated and in the real system correspond to the observed
pair of high and low frequency magneto- and electro-active modes
(Fig. \ref{fmodes}). Secondly, one phason mode is expected for the
oscillations of the antiferromagnetic vector within the bc-plane.
Without bc-plane anisotropy the frequency of this mode is zero
(Goldstone mode). The phason mode correspond to the AFMR with
excitation condition $\tilde{h}\|a$ which is similar to the F-mode
\cite{lsmo} in a canted antiferromagnet.

\begin{figure}[]
\includegraphics[width=8cm,clip]{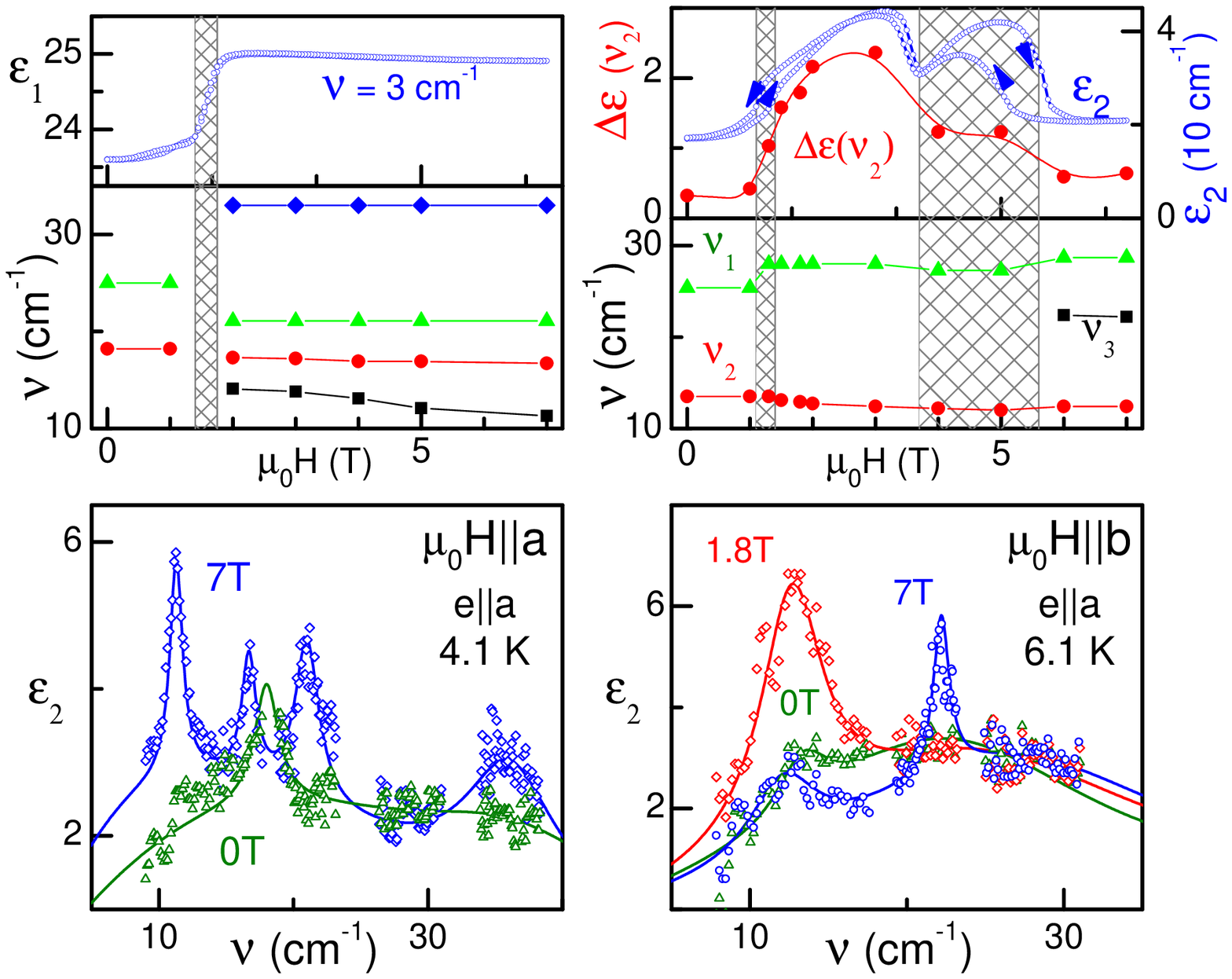}
\caption{(color online) Behavior of the electromagnons in TbMnO$_3$
in external magnetic fields. Left panels: $\mu_0H \|a$, right panels
$\mu_0H \|b$. Lower panels - examples of the absorption spectra.
Lines are fits using Lorentzian line shape. Upper panels - field
dependence of the mode frequencies and of the dielectric constant.
$\Delta \varepsilon$ represents the dielectric contribution of the
$\nu_2$ mode. Lines are guides to the eye.} \label{fb}
\end{figure}

We discuss now the behavior of the electromagnons in TbMnO$_3$ in
external magnetic fields parallel to the crystallographic a- and
b-axes. These results are represented in Fig. \ref{fb}. Slight
deviation of the shape of the modes from that published previously
\cite{nphys,review} is basically due to weak sample dependence of
the spectra. We recall that according to the previous experiments,
the external fields along the c-axis suppress the electromagnons
\cite{nphys} and induce a canted antiferromagnetic structure
\cite{kimura05,argyriou}. External magnetic field along the a- and
b- axes leads to the rotation of the magnetic cycloid from the
bc-plane to the ac-plane. Applying magnetic fields $\mu_0H
> 5$ T along the b-axis allows for complete rotation of the cycloid
plane. Along the a-axis fields of more than 10 T are necessary and
only a tilting of the cycloid can be achieved using our magnet. In
both cases and already for fields above $\approx 2$ T substantial
changes in the spectral structure of the electromagnons can be
observed. For $\mu_0H\|a$ instead of initially two electromagnons we
observe four new modes in the spectra. We attribute the appearance
of these modes to the change of the excitation conditions of the
magnetic cycloid due to tilting. For magnetic fields along the
b-axis and in low fields a redistribution of the spectral weight of
electromagnons is seen. After the transition to the ab-cycloid an
additional mode appears at 22 \cm. Qualitatively similar complexity
of the magnetic modes has been observed recently in neutron
scattering experiments \cite{braden3} and complicated excitation
conditions have been determined. However, in our case all observed
modes retain the excitation condition $\tilde{e} \| a$.

\begin{figure}[]
\includegraphics[width=5cm,clip]{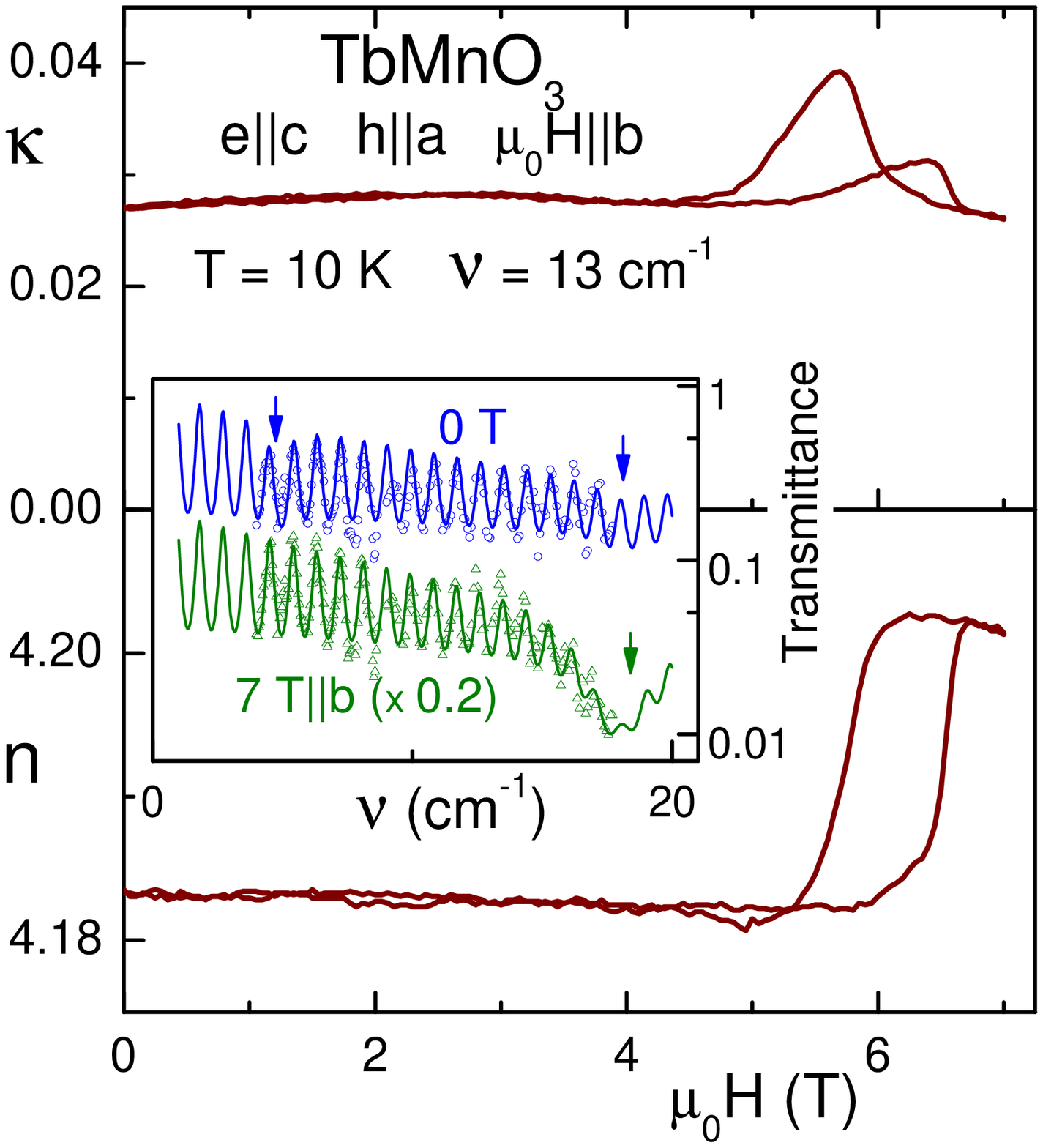}
\caption{(color online) Magnetic field dependence of the refractive
index in TbMnO$_3$ along the c-axis demonstrating the absence of the
electromagnons in c-direction. Close to $\mu_0H = 6$ T the magnetic
bc-cycloid is switched to the ab-cycloid. According to the simple
arguments, the electromagnon contribution would expected to rotate
from a-axis to the c-axis (see text). The inset shows examples of
the transmittance spectra. Here the spectra for $\mu_0H=7$ T have
been shifted for clarity.} \label{fc}
\end{figure}

From the simple arguments the rotation of the magnetic cycloid from
bc-plane to the ac-plane should simultaneously switch the excitation
conditions for the electromagnons from $\tilde{e} \| a$ to
$\tilde{e} \| c$. In order to check this prediction, a series of
transmittance experiments for ac-electric fields along the c-axis
has been carried out. Typical result of these experiments is shown
in Fig. \ref{fc}. The data are given in the representation
$n+i\kappa = \sqrt{\varepsilon \mu}$ because electric and magnetic
contributions are mixed in these experimental geometry. However, the
main result can be stated already at this point: All changes
detected along the c-axis as function of magnetic field are
extremely weak and electromagnons are not observed along the c-axis.
Indeed the measured changes in the refractive index at the magnetic
transition amount roughly 0.5\%. This value should be compared with
the dielectric strength of the electromagnons along the a-axis which
reaches $\approx 5 \%$. However, even the observed small changes
along the c-axis are not due to magnetoelectric contribution but due
to shifts of the AFMR frequencies, i.e. are of purely magnetic
origin. The changes in the refractive index, shown in Fig. \ref{fc}
can be well explained using increased intensity of the AFMR mode at
$\sim 18$ \cm, which is excited by $\tilde{h} \| a$ in this
experimental geometry (Fig. \ref{ftranb0} and inset in Fig.
\ref{fc}). This result is challenging for the interpretation of the
electromagnons as spin excitations of spiral structures. In addition
to above mentioned two-magnon scenario an explanation on the basis
of the DM coupling has been suggested recently \cite{mostovoy2}.
Within this model structural peculiarities of perovskite manganites
are responsible for exclusive $\tilde{e} \| a$ excitation conditions
for the electromagnons.

In conclusion, we have carried out terahertz experiments in
multiferroic TbMnO$_3$ in order to study magnetic and
magnetoelectric excitations in this compound. Using different
experimental geometries it was possible to separate magnetoelectric
excitations (electromagnons) from antiferromagnetic resonances. In
agreement with the neutron scattering data we observe the splitting
of the electromagnon in the cycloidal phase. The frequencies of two
AFMR modes coincide with the electromagnon frequencies. This
indicates that both excitations correspond to the same mode of the
magnetic cycloid and the magnetoelectric excitations can be excited
via magnetic as well as via electric fields. In external magnetic
fields an increased complexity of the magnetoelectric excitations
can be observed. No electromagnon contribution can be detected along
the c-axis even after the induced rotation of the magnetic cycloid
from bc- to ab-plane. This clearly contradicts simple explanation of
the electromagnons based on the magnetic cycloid.

We thank Anna Pimenov for help in magnetic and X-ray measurements.
This work has been supported by DFG (PI372, SFB484) and by RFBR
(06-02-17514, 09-02-01355).


\begin{thebibliography}{99}

\bibitem{fiebig} M. Fiebig, J. Phys. D: Appl. Phys. \textbf{38}, R123
(2005).

\bibitem{tokura} Y. Tokura, Science \textbf{312}, 1481 (2006).

\bibitem{cheong} S.-W. Cheong and M. Mostovoy, Nature Mater. \textbf{6}, 13(2007).

\bibitem {khomskii} D. Khomskii, J. Magn. Magn. Mater. \textbf{306}, 1
(2006).

\bibitem{ramesh} R. Ramesh  and N. A. Spaldin, Nature Mater. \textbf{6}, 21
(2007).

\bibitem{kimura} T. Kimura \textit{et al.}, 
Nature  \textbf{426}, 55 (2003).

\bibitem{goto} T. Goto \textit{et al.}, 
Phys. Rev. Lett. \textbf{92}, 257201 (2004).

\bibitem{hur} N. Hur \textit{et al.}, 
Nature  \textbf{429}, 392 (2004).


\bibitem{nphys} A. Pimenov \textit{et al.}, 
Nature Physics \textbf{2}, 97 (2006).

\bibitem{sushkov} A. B. Sushkov \textit{et al.}, 
Phys. Rev. Lett. \textbf{98}, 027202 (2007).

\bibitem{euymno} A. Pimenov \textit{et al.}, 
Phys. Rev. B \textbf{77}, 014438 (2008).

\bibitem{aguilar} R. V. Aguilar \textit{et al.}, 
Phys. Rev. B \textbf{76}, 060404(R) (2007).

\bibitem{kida} N. Kida \textit{et al.}, 
Phys. Rev. B \textbf{78},104414 (2008).

\bibitem{braden} D. Senff \textit{et al.}, 
Phys. Rev. Lett. \textbf{98}, 137206 (2007).

\bibitem{braden2} D. Senff \textit{et al.}, 
J. Phys.: Cond. Matter \textbf{20} , 434212 (2008).

\bibitem{review} A. Pimenov \textit{et al.}, 
J. Phys.: Cond. Matt. \textbf{20} 434209 (2008).

\bibitem{chupis} V. G. Baryakhtar and I. E. Chupis,  Sov. Phys.-Sol. State \textbf{11}, 2628 (1970).

\bibitem{chupis2} I. E. Chupis, Low Temp. Phys. \textbf{33}, 952 (2007).

\bibitem{mostovoy} M. Mostovoy,
Phys. Rev. Lett. \textbf{96}, 067601 (2006).

\bibitem{harris} A. B. Harris, Phys. Rev. B \textbf{76}, 054447 (2007)

\bibitem{sousa} R. de Sousa and J. E. Moore, Phys. Rev. B \textbf{77}, 012406
(2008).

\bibitem{katsura05} H. Katsura, N. Nagaosa, and A. V. Balatsky, Phys. Rev. Lett. \textbf{95}, 057205
(2005).

\bibitem{katsura} H. Katsura, A. V. Balatsky, and N. Nagaosa,
Phys. Rev. Lett. \textbf{98}, 027203 (2007).

\bibitem{sergienko} I. A. Sergienko and E. Dagotto,
Phys. Rev. B \textbf{73}, 094434 (2006).

\bibitem{sergienko2} I. A. Sergienko, C. \c{S}en, and E. Dagotto, Phys. Rev. Lett. \textbf{97}, 227204
(2006).

\bibitem{sushkov2} A. B. Sushkov \textit{et al.}, 
J. Phys.: Cond. Matt. \textbf{20} 434210 (2008)

\bibitem{kida2} Y. Takahashi \textit{et al.}, 
Phys. Rev. Lett.
\textbf{101}, 187201 (2008).

\bibitem{kida3} N. Kida \textit{et al.}, J. Phys. Soc. Jpn. \textbf{77},
123704 (2008).

\bibitem{kimura05} T. Kimura \textit{et al.}, 
Phys. Rev. B \textbf{71}, 224425
(2005).

\bibitem{quezel} S. Quezel \textit{et al.}, 
Physica B\&C
\textbf{86}, 916 (1977).

\bibitem{kajimoto} R. Kajimoto \textit{et al.}, 
Phys. Rev. B \textbf{70}, 012401 (2004).

\bibitem{kenzelmann} M. Kenzelmann \textit{et al.}, 
 Phys. Rev. Lett. \textbf{95}, 087206 (2005).

\bibitem{argyriou} D. N. Argyriou \textit{et al.}, Phys. Rev. B \textbf{75},
020101 (2007).

\bibitem{strempfer} J. Strempfer \textit{et al.}, 
Phys. Rev. B \textbf{78}, 024429 (2008).

\bibitem{braden3} D. Senff \textit{et al.}, Phys. Rev. B \textbf{77}, 174419
(2008).

\bibitem{volkov} A. A. Volkov \textit{et al.}, 
Infrared Phys. {\bf 25}, 369 (1985);
A. Pimenov \textit{et al.}, 
Phys. Rev. B \textbf{72}, 035131 (2005).

\bibitem{lsmo} D. Ivannikov \textit{et al.}, 
Phys. Rev. B \textbf{65}, 214422 (2002).

\bibitem{mostovoy2} R. V. Aguilar \textit{et al.}, arXiv:0811.2966
(unpublished).



















\end{thebibliography}
\end{document}